\documentclass[journal]{IEEEtran}

\ifCLASSINFOpdf
   \usepackage[pdftex]{graphicx} 
\else
   \usepackage[dvips]{graphicx}
   
  \graphicspath{{../eps/}}
  
\fi

\usepackage{dblfloatfix}

\usepackage[cmex10]{amsmath}
\usepackage{caption}
\usepackage{float}

\usepackage[table]{xcolor}
\usepackage{multirow}
\usepackage{multicol}
\usepackage{cuted}

\usepackage{enumerate}
\usepackage{subfigure} 
\usepackage{acronym}
\usepackage{epstopdf}
\usepackage{cite}
\usepackage{xcolor}
\usepackage{fancyhdr}

\usepackage{amsfonts}   

\usepackage{gensymb}
\usepackage{easyReview}
\usepackage[binary-units]{siunitx}
\def\BibTeX{{\rm B\kern-.05em{\sc i\kern-.025em b}\kern-.08em T\kern-.1667em\lower.7ex\hbox{E}\kern-.125emX}}

\input{AcronymList.tex}
\begin{document}
\bstctlcite{IEEEexample:BSTcontrol}
\captionsetup[figure]{labelfont={default},labelformat={default},labelsep=period,name={Figure}}
%
\title{Generalized Radio Environment Monitoring for Next Generation Wireless Networks}
\author{Halise~T{\"u}rkmen,~Muhammad~Sohaib~J.~Solaija,~\IEEEmembership{Student Member, IEEE},\\~Haji M.~Furqan~and~H{\"u}seyin~Arslan,~\IEEEmembership{Fellow,~IEEE}
\thanks{Authors are with Istanbul Medipol University. H. Arslan is also with University of South Florida.}

\thanks{This work has been submitted to the IEEE for possible publication. Copyright may be transferred without notice, after which this version may no longer be accessible.}}
\maketitle

\begin{abstract}
Enabling technologies of 5G and beyond wireless communication networks, such as millimeter-wave communication, beamforming, and multiple-input multiple-output (MIMO) antenna systems, are becoming increasingly dependent on accurate information of the physical environment for optimized communication performance. The acquisition of this information is a significant challenge, which can be eased by knowledge of the radio environment. Radio environment mapping (REM) has long been considered an enabler of cognitive radios (CRs). However, the limitations of radios effectively confined the application of REM to spectrum sensing and interference mapping. With the advent of more capable software-defined radios and advanced networks, the idea of a truly intelligent communication system empowered by enhanced REM can be realized. To this effect, we propose the generalized radio environment monitoring (G-REM) concept to include all aspects of the radio environment. We outline a general, self contained framework which enables radio environment monitoring in standalone devices. The presented G-REM framework comprises of different information sources, sensing methods, sensing modes, and mapping techniques to enable the realization of more reliable, secure, efficient, and faster future wireless networks. The accompanying challenges and future research direction are also provided.
\end{abstract}


\vspace{-3.5pt}
\section*{Introduction}
\label{Sec:Introduction}
While the \ac{5G} of wireless networks already diversified the communication paradigm by introducing services such as \ac{mMTC} and \ac{uRLLC} in addition to the evolving \ac{eMBB} \cite{series2015imt}, this trend of enriching the application domain is set to continue with the upcoming \ac{6G}. Applications such as \ac{LDHMC}, \ac{eLPC}, \ac{XR}, and previously unexplored combinations of the \ac{5G} services are expected to dominate the future networks \cite{gui20206g}.

The diversity in requirements necessitates networks capable of intelligently adapting to the varying user requirements and network provisions. The efforts to impart intelligence to communication systems in the form \acp{SON} \cite{3GPP_36_902} and \acp{CR} \cite{mitola1999cognitive} have fallen significantly short of the said goal. \ac {SON} primarily focuses on observation of measurements within the network to configure its parameters for easier configuration, optimization, and management while the prevalent \ac{CR} concept is limited to monitoring the radio spectrum for opportunistic access.

Neither of the aforementioned mechanisms, however, considers/exploits the knowledge of other parameters such as user-related information, waveform information, and propagation environment. This can be leveraged not only to improve the communication performance by empowering technologies such as \ac{mmWave} communications, \ac{MIMO} systems, \ac{CoMP} networks, and multinumerology systems - all of which depend on the radio environment - by providing information related to the propagation conditions, but also enhance the sensing capabilities for applications such as motion detection, gesture recognition, health monitoring, and so on. With the exception of some isolated efforts to enable easier construction of local and cached \acp{REM}, this task has been primarily left to \acp{BS} and network control centers due to limited processing, storage, and communication capabilities of legacy devices. This has meant the application of \ac{REM} in real life is limited to \ac{RSSI} mapping and spectrum monitoring for interference mitigation and \acp{CR}, respectively.

\begin{figure*}[h]
    \centering
    \includegraphics[width=1\textwidth]{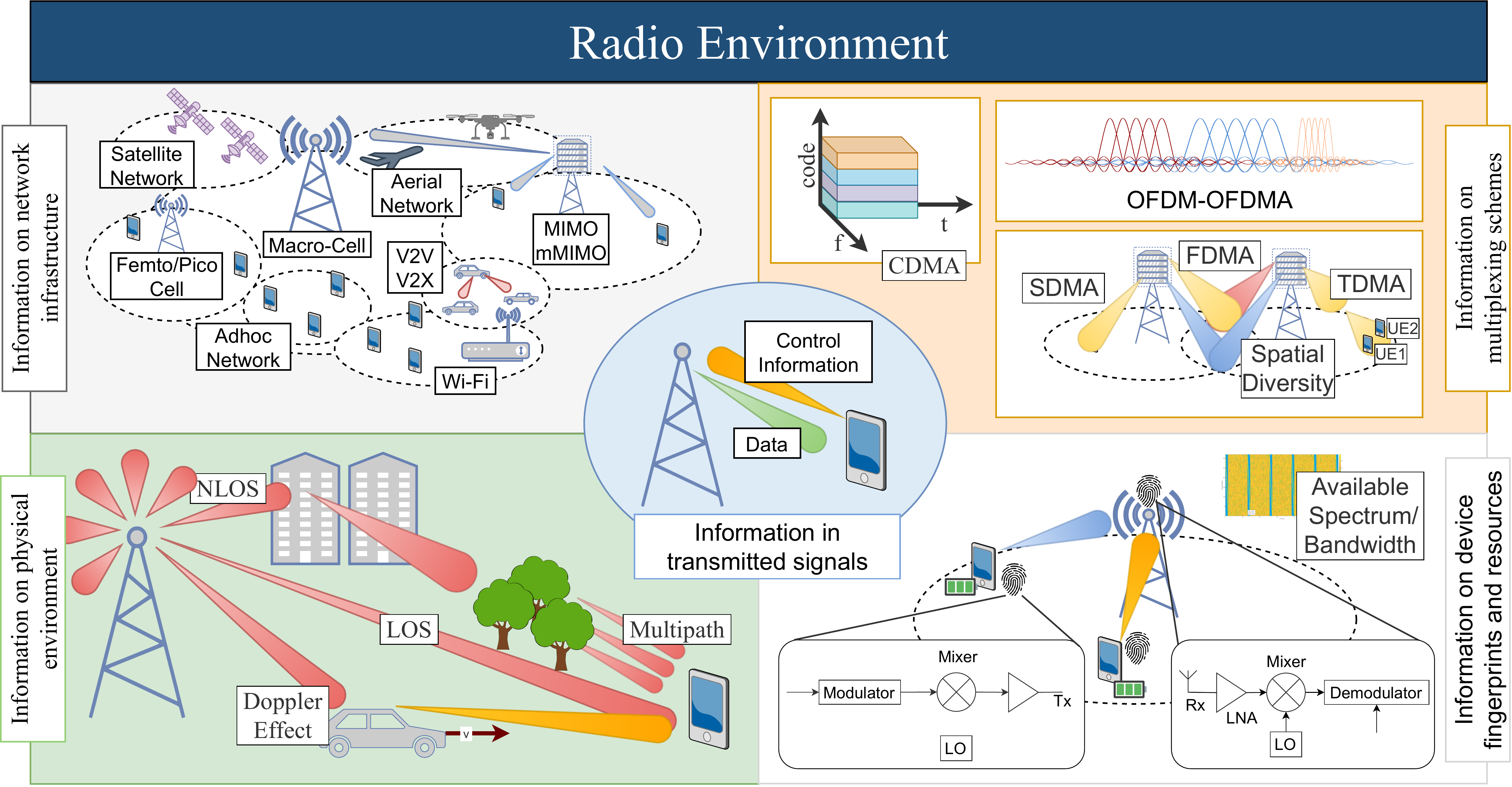}
    \caption{Illustration of the generalized radio environment monitoring (G-REM) concept. The radio environment information is made up of (but not limited to) knowledge regarding network infrastructure, propagation environment, attributes of the physical signal (multiple accessing scheme, waveform, modulation, etc.), and characteristics of the user equipment.}
    \label{fig:radio_environment}
\end{figure*}

However, the greater availability of more capable hardware and wireless technologies, such as fog/edge computing, high capacity backhaul links, and network slicing, combined with emerging integrated sensing and communication technologies, means more devices can construct their own \ac{REM}. This motivates the concept of a \ac{GREM}, presented in this work, targeted at acquiring multi-dimensional information of the radio environment. It includes information of the network, environment, \ac{LoS} characteristics, multiple accessing techniques, waveform, hardware impairments, and so on, as illustrated in Fig. \ref{fig:radio_environment}. The environmental information can be leveraged to optimize the deployment of \acp{RIS} and relays to improve communication quality, mobility patterns can be utilized to perform predictive handovers ensuring ubiquitous connectivity, while user behavior and preferences can be employed to improve content caching alleviating backhaul load. Moreover, different sensing applications such as gesture recognition, localization, and user tracking can also be enabled by improved environment monitoring.

To this end, we contribute the following to the literature:
\begin{itemize}
    \item The \ac{GREM} framework is presented which covers the various construction methods and different parameters of interest that can be learnt regarding a radio environment.
    \item The applications of \ac{GREM} in the context of future wireless networks are discussed including the provision of secure communication and optimized network deployment.
    \item The challenges in the way of realizing the proposed \ac{GREM} concept are highlighted and directions for possible solutions are proposed.
\end{itemize}
The structure of the paper is as follows. The general framework of \ac{REM} is described, providing both its constituents and construction methods. Applications of this framework in the context of next generation wireless networks are then discussed before the research challenges and future directions are highlighted. In the end, conclusions are drawn.

\vspace{-5pt}
\section*{Generalized Radio Environment Monitoring Framework}
\label{sec:framework}

\begin{figure*}[ht]
    \centering
    \includegraphics[scale = 0.9]{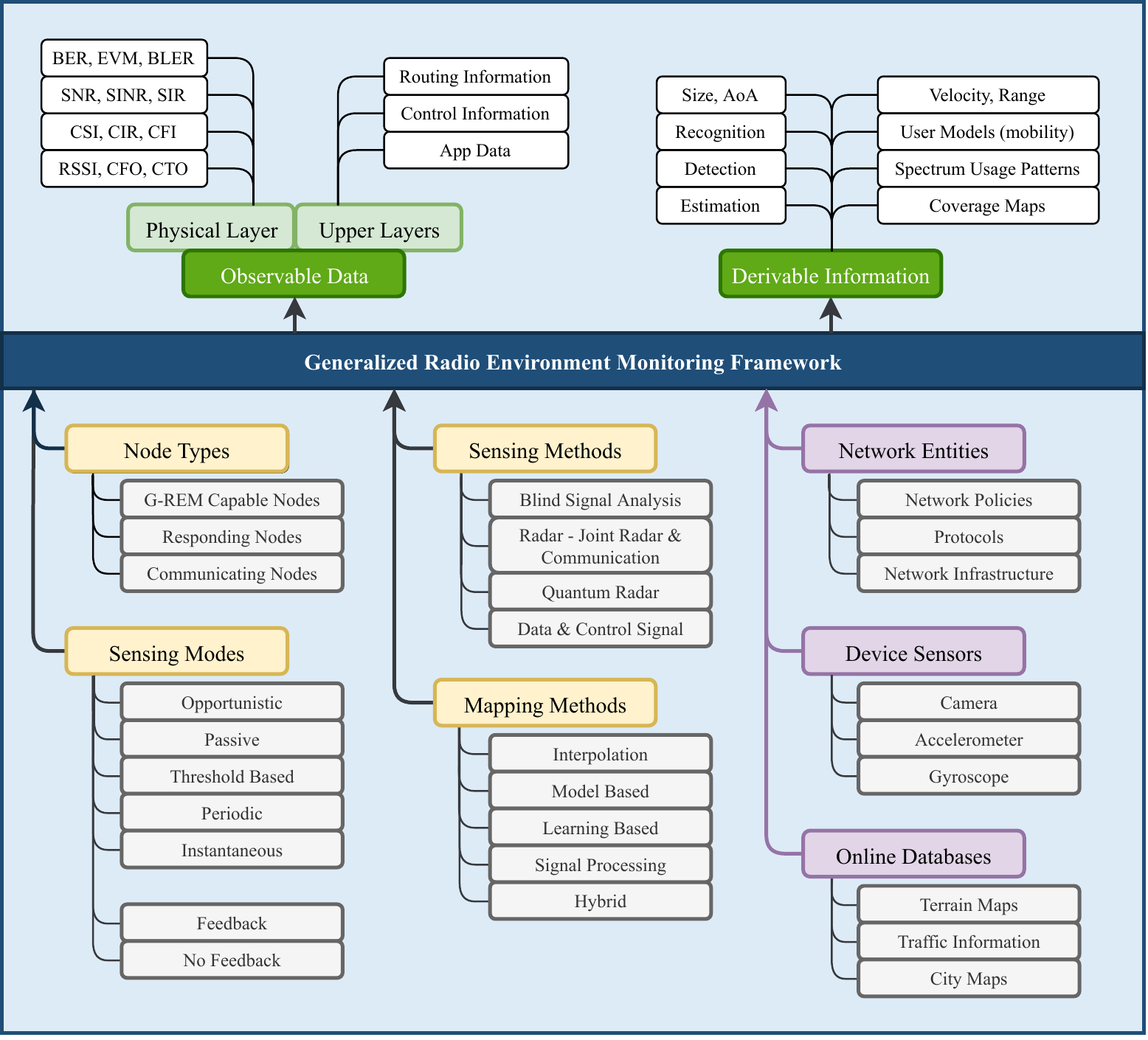}
    \caption{Generalized radio environment monitoring framework}
    \label{fig:rem_framework}
\end{figure*}

\subsection*{Preliminaries}
There is a need to clarify some terminologies that are frequently used interchangeably in the literature:
\begin{itemize}
    \item Radio Environment Sensing: The process and methods of extracting or measuring raw data from the received signals.
    \item Radio Environment Map/Mapping: A collection of integrated databases containing information relevant to the radio environment, and the data processing/mapping methods \cite{zhao2006network} required to map sensed data to other domains or extract user/environment patterns.
\end{itemize}

In addition to these, low level data is the raw sensed data without any processing, also referred to as \emph{observable} data, whereas high level or \emph{derivable} information, is a result of intelligent interpretation of the low level data. Another distinction to be kept in mind is between sensing and communication signals. The former do not contain user information and have better sensing capabilities compared to the latter, while the latter contain user data and even though they can be enhanced for sensing, that is not their sole purpose. The rest of the article will assume the aforementioned definitions for these terms. 

\vspace{-7pt}
\subsection*{Overview of the Framework}
\par{The framework comprises of the various processes associated with the initiation and management of environment sensing, collection of data, and storage or processing this data to gain \emph{radio environment awareness}, as illustrated in Fig. \ref{fig:rem_framework}. The different entities of the \ac{GREM} framework are:}
\begin{itemize}
    \item nodes participating in the radio environment monitoring process
    \item methods to perform radio environment sensing,
    \item sensing modes enabling the different sensing methods as well as sensing qualities
    \item methods to map the sensed spectrum and environment measurements to information in different domains
    \item external sources of information to aid in the sensing process
    \item the sensed data and the derived information
\end{itemize}

\vspace{-6pt}
\subsection*{\ac{GREM} Node Types}
\par{Nodes are taken to be wireless-capable devices in the network or area of sensing. Depending on their role in the radio environment sensing process, they are classified into three groups:
\begin{itemize}
    \item \ac{GREM} Capable Nodes: These nodes can initiate a sensing or monitoring instance in response to information requests by applications and have the resources to perform the required sensing and/or mapping methods.
    \item Responding Nodes: These nodes actively take part in the sensing instance by either transmitting sensing signals or their measurements from received sensing signals.
    \item Communicating Nodes: These nodes are not active participants of the sensing instance and are transmitting communication signals only.
\end{itemize}}

\begin{figure*}[ht]
    \centering
    \includegraphics[scale = 0.95]{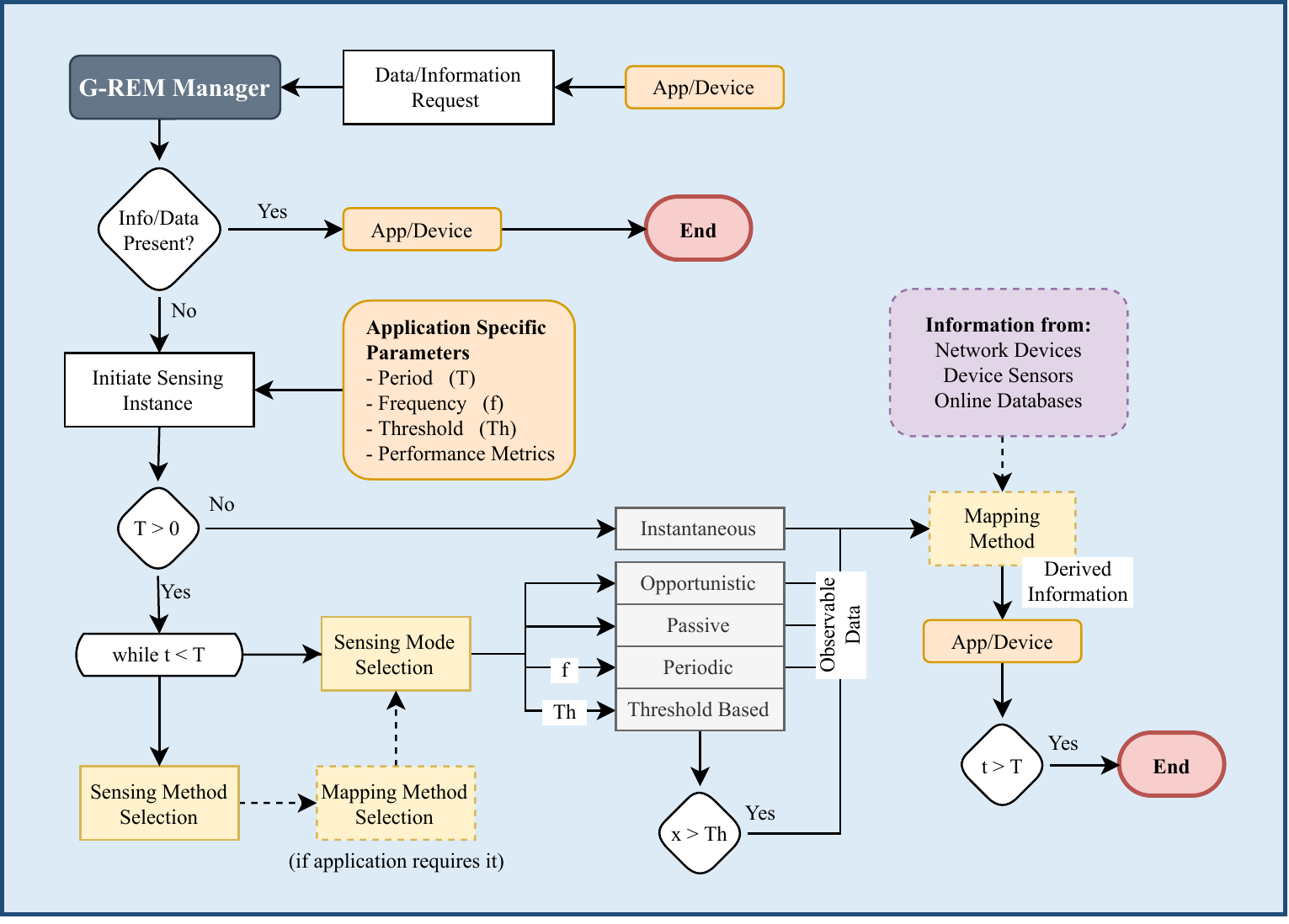}
    \caption{A schematic depicting a simple \ac{GREM} process}
    \label{fig:grem_process}
\end{figure*}

\vspace{-10pt}
\subsection*{Sensing Modes}
\vspace{-2pt}
\par{Before going to the specific sensing modes, it is important to define a \textit{sensing instance} as the duration when the \ac{GREM} capable device actively monitors the radio environment. This state is induced by an information request from an application and lasts for a fixed amount of time. Choosing the appropriate mode for any sensing instance and its parameters is crucial in many aspects. 
\begin{itemize}
    \vspace{+5pt}
    \item Opportunistic: The \ac{GREM} capable node uses its transmitted and received communication signals for radio environment monitoring. 
    \vspace{+5pt}
    \item Passive: The \ac{GREM} capable node does not take part in any communication or active sensing process, but exploits transmissions between other devices.
    \item Periodic: Sensing signals are transmitted with a certain frequency, usually related to the required resolution or accuracy.
    \item Instantaneous: Here, a single burst of measurements is taken. This mode can be used if there is already prior knowledge of the environment and a validation check is required.
    \item Threshold Based: Sensing occurs as described in one of the above modes, but mapping or forwarding information only occurs if the raw data passes a certain threshold. 
\end{itemize}
Depending on the sensing process, these modes can be used in two configurations: Feedback and No Feedback. In the Feedback configuration, the \ac{GREM} capable node transmits sensing signals and responding nodes perform radio environment sensing. Then, they forward their measurements to the \ac{GREM} capable node. In No Feedback configuration, the responding nodes transmit the sensing signals and the \ac{GREM} capable node itself performs sensing measurements.}

\par{The sensing mode is mostly dependent on the device state, sensing, and mapping methods and sensing parameters are dependent on the communication traffic and available resources. Each mode has its own advantages and disadvantages regarding power and spectral efficiency, accuracy, resolution, and communication performance. For example, opportunistic, passive, and instantaneous sensing modes add little to no additional traffic load, whereas it can be quite high for periodic sensing mode. However, if there are insufficient communication transmissions and/or the application requires continuous measurements, the first three sensing modes are not suitable and this load is necessary.}

\vspace{-9pt}
\subsection*{Sensing Methods}
\par{Sensing methods, the most popular of which is \emph{\ac{radar}}, focus on the acquisition of low level data. The \ac{radar} works by sending transmissions, usually a linear frequency modulated continuous chirp, and correlating the reflected signals with the transmitted ones to get the delay and Doppler frequency. These are then mapped to the range and velocity of the detected object. The angle of arrival (and therefore, the position) can also be detected using rotating directional antennas or phased-array antennas. The size of the object can be known through successive, fine beamed measurements or reflected power \cite{bassemrdar}.}
\par{The relatively new \emph{\ac{JRC}} concept applies radar principles to communication systems. Three main concepts are studied here: coexistence, RadComm systems, and CommRad systems. Coexistence refers to radar and communication waveforms utilizing the same resources while being orthogonal in at least one domain. RadComm and CommRad systems are also referred to as dual function radar communication (DFRC) systems. The main idea is to develop hybrid waveforms with both radar and communication capabilities. In RadComm systems, radar waveforms are modulated with data to give it communication capabilities. CommRad systems have communication waveforms with enhanced correlation properties to improve radar performance, generally through using highly correlated sequences or multiplexing radar and communication waveforms.}
\par{The \emph{data and control signals} are also a source of information. The communication related methods in the \ac{PHY} and \ac{MAC} layer, such equalization which gives \ac{CSI} as a byproduct, result in many of the low level measurements. The data, if it can be demodulated, can also contain information that can be used to infer high level data and improve communication. Control signals contain information on devices, spectrum occupancy, multiplexing methods, and so on. \cite{3GPP_38_331}. Lastly, \ac{BSA} is another emerging concept in wireless communication where signal detection and analysis is performed with limited, global information, and machine learning methods. Though the sensing methods may differ, the observable data remains more or less the same.}

\vspace{-7pt}
\subsection*{Mapping Methods}
\par{The mapping methods, shown in Fig. \ref{fig:rem_framework}, can be functions, models, or algorithms used to relate data from one domain to another or map the observed data to the desired information. For example, various interpolation algorithms can be used to estimate missing values. Signal processing and transformation algorithms are used to filter noise and erroneous measurements and analyze the data in different domains \cite{ma2019csisensingsurv}.}

\vspace{-5pt}
\subsection*{External Sources of Information}
\par{Non-radio related information can also be used to improve the accuracy and performance of \ac{GREM}. There are three main sources of external information.}

\subsubsection{Network Entities}
\par{This refers to the elements of the network infrastructure (e.g. \acp{BS}, network control centers, access points and routers). Information provided by these sources could be network policies/protocols, network devices and users, devices/users of other operators, and so on. }

\subsubsection{Device Sensors}
\par{This can be any sensor(s) on the device (e.g. accelerometer, gyroscope, camera, pressure sensor, etc.) that can give information on relative velocity, orientation, surroundings, and so on.}

\subsubsection{Online Databases}
\par{This information can be accessed through the internet. Terrain maps, traffic patterns, and so on are examples of possible information.}

\vspace{-7.5pt}
\subsection*{\ac{GREM} Process}
\par{The general operation of the \ac{GREM} framework is given in Fig. \ref{fig:grem_process}. An information or continuous monitoring request by a device or application is made to the \ac{GREM} manager. The lack of requested information causes the initiation of a sensing instance. Here, certain parameters may be provided by the application to determine the sensing mode, sensing method, and mapping method. Example parameters can be:
\begin{itemize}
    \item Period \(T\): Duration of the sensing instance. 
    \item Frequency \(f\): Number of sensing transmissions per second. It is related to the time resolution of the derived information.
    \item Threshold \(Th\): Cutoff value for the threshold based sensing mode.
    \item Performance Metrics: Defines the desired quality of the sensed data and derived information, the sensing/mapping methods, and sensing signal parameters like bandwidth.
\end{itemize}
The parameters must be used to perfectly meet the sensing requirements of the applications. However, since the real-life scenarios may not permit this, optimization must be performed by the framework.}
\par{Once the sensing mode is selected, the radio environment sensing takes place and the observed data is processed using the radio environment mapping methods, if it is required, and sent to the application. When the sensing period is over, the sensing instance is terminated. The sensing period may be indefinite for continuous monitoring applications, like fall detection, or fixed for applications with a known duration, like beam management for communication. }

\begin{figure*}[ht]
    \centering
    \includegraphics[scale=0.9]{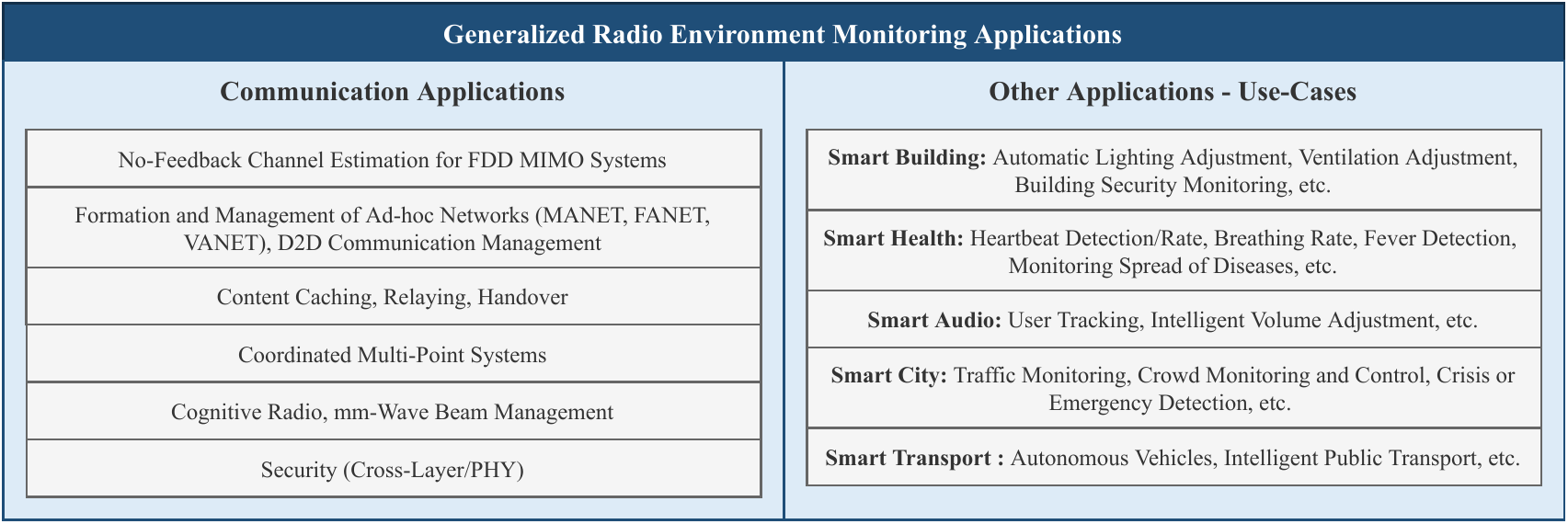}
        \caption{Some applications of the \ac{GREM} framework in communication and other fields}
    \label{fig:gremapps}
\end{figure*}

\vspace{-7pt}
\section*{Applications of \ac{GREM}}
\label{sec:applications}

\subsection*{Enhancing Physical Layer Security}
The proposed \ac{GREM} framework promises to be a critical enabler of the cognitive security concept, capable of providing adaptive security to users depending on application requirements, network conditions, and expected threats. Information from the network, particularly \ac{5QI}, defines the application requirements \cite{3GPP_23_501} while other \ac{GREM} parameters such as hardware impairments allow user authentication, and location information can be used to determine threat level (as in the case of vehicular communication) or empower directional transmission to ensure secure transmission \cite{liu2016physical}.

\vspace{-11pt}
\subsection*{\ac{GREM}-aided CoMP Systems}
\Ac{CoMP} systems utilize the coordination between geographically separated transmission points to improve communication performance, particularly at cell edges. \ac{GREM} based information regarding interference levels, propagation conditions, network congestion, and possible mobility can help identify the best possible coordination clusters for users, leading to better \ac{QoS}.

\vspace{-12pt}
\subsection*{Aiding Intelligent Multi-Antenna Systems}
\par{\ac{GREM} can provide information on the spectrum and predict channel quality variation in time using spectrum usage patterns, mobility information, interference patterns, and much more. It can also provide information on the position and trajectory of the \ac{UE}, as well as obstacles, facilitating dynamic beam-management for \ac{mmWave} \ac{MIMO} systems and optimizing \ac{RIS} placement and control.}

\par{One obstacle in the mass deployment of \ac{mmWave} systems is the channel estimation overhead. For this reason, these systems are primarily \ac{TDD} systems, where the channel is assumed to be reciprocal and has a geometric correlation. However, recent works in the literature have been able to estimate some channel parameters in the down-link for \ac{FDD} \ac{mMIMO} systems using the spatial reciprocity in the propagation paths of the wireless signals. They map the channel information of the uplink channel to the downlink channel by reconstructing the multipath \cite{dina}. This is a direct application of the \ac{GREM} framework.}

\vspace{-12pt}
\subsection*{Forming and Coordinating Ad-hoc Networks}
\par{With the steady increase of different devices and their communication requirements - from sensor networks to \ac{IoT} devices to \ac{V2X} communications, the under-realized ad-hoc network concept needs to be revisited. The \ac{GREM} framework can improve the cooperation and coordination in these networks. Devices performing the \ac{GREM} process can behave intelligently. For example, \ac{GREM} can be used to differentiate among the devices and form social patterns. In this way, reputation based communication and security methods can be implemented. Mobility patterns of the devices can be used to dynamically trim or expand the network by helping to decide which devices have left the area and which are temporarily out of reach. Intelligent content caching can also be performed by selecting the devices likely to remain in the network longest for content hosting.}

\vspace{-12pt}
\subsection*{Increasing Reliability of Handover Schemes}
\par{Handover is still a big issue in cellular systems. The \ac{GREM} framework can be used by the \acp{BS} to extract mobility patterns and associate \acp{UE} to them. Similarly, \ac{GREM} capable \acp{UE} can construct their specific patterns and inform the \ac{BS}. In this way, \ac{UE} traveling along cell edges can be differentiated from \ac{UE} leaving the cells and the handover protocols can be initiated accordingly.}

\vspace{-11pt}
\subsection*{Realizing True Flexibility for 6G Visions}
\par{The flexibility required in 5G systems was achieved by the band-aid approach of multiple numerologies. However, 6G visions have higher demands in terms of flexibility, with the aim of not only serving even more devices but serving these devices \emph{better} - with even lower latency, more throughout, more reliability, and security. Therefore, not only do 6G networks have to be incredibly flexible, they have to be proactive as well. In order to enable this, any and all information on the radio environment is crucial, which is where \ac{GREM} comes in. \ac{GREM} provides radio environment awareness and contains methods to make future predictions. These resources can aid cognition and true flexibility in 6G devices.  } 

\vspace{-5pt}
\section*{Challenges and Future Directions}
\label{sec:challenges}
\subsection*{Securing \ac{GREM}}
The above-mentioned applications of \ac{GREM} highlight its importance in ensuring smarter, more reliable, and secure future networks. These networks can be used for communication, sensing, and so on. Considering the extent of information, related to the users, network, and radio environment that resides in a \ac{GREM}, its security becomes a major concern. The knowledge of \ac{GREM} information, if accessed by an illegitimate node, can be used to disrupt or manipulate the system. In terms of communication, it can lead to poor data transmission while other applications such as sensing can be misled to make poor (activity/gesture/mobility) detection. Ensuring the privacy, reliability, and authenticity of all the aspects \ac{GREM} - including sensing modes, information, and processing - is a challenging concern that needs to be addressed.

\vspace{-10pt}
\subsection*{Information Freshness vs. Overhead Trade-off}
The extracted radio environment information is directly proportional to the incurred overhead. This overhead is manifested in terms of additional transmissions, information exchange between nodes, processing for information extraction from the physical signal, and storage of the acquired information. Analysis to determine the feasibility of such a comprehensive \ac{GREM} against the gains provided in terms of communication enhancement needs to be carried out.

\vspace{-5pt}
\subsection*{Sensing Scheduling and Coexistence}
\par{Wireless networks are first and foremost a means of communication between devices. As such, the main priority is communication and any sensing activity must not deteriorate the communication performance. However, maintaining high sensing quality without causing degradation to the communication performance is difficult. The scheduling of periodic sensing signals in high communication traffic areas and times is a serious problem which is compounded by the fact that communication transmissions are relatively random and may have low latency requirements. Depending on the application, reduced sensing quality can also be fatal (e.g. in autonomous vehicles or health monitoring systems). Furthermore, in sensing aided communication applications, like beamforming where beam alignment is dependent on information about the obstacles, optimizing sensing and communication performances can be a conundrum.}

\vspace{-5pt}
\subsection*{\ac{GREM} for Green Networks}
Energy efficiency is a critical application for applications/services such as \ac{IoT} and \ac{mMTC}. The general approaches from the network perspective in this regard include energy harvesting or \ac{SWIPT} and \ac{BS} sleeping \cite{buzzi2016survey}. Future realizations of \ac{REM} need to have an additional layer comprising of information that helps in improving energy efficiency, such as optimized beamforming vectors for energy receivers.

\vspace{-5pt}
\subsection*{Defining \ac{GREM} Quality Metrics}
The conventional \ac{REM} uses performance metrics such as \textit{missed detection} or \textit{false alarm} rates \cite{yilmaz2013radio}. However, such metrics are limited to binary classification problems such as spectrum occupancy. However, given that \ac{GREM} has outputs that can be continuous in nature, new performance metrics need to be defined. Similar to the aforementioned metrics, quality indicators related to regression problems can be borrowed from the domain of machine learning.

\vspace{-5pt}
\subsection*{Integration of \ac{GREM} and Novel Technologies}
\vspace{-2pt}
\par{Wireless communication is a rapidly developing area. However, every new development comes with its own complications which must be resolved before the communication system can benefit from these technologies. \ac{RIS} is one such technology currently being evaluated for integration to the communication framework. It must also be evaluated with respect to \ac{GREM}. For example, given that they are passive devices, how can the \ac{GREM} framework differentiate between channel due to \ac{RIS} and its phase changes and other objects?}
\par{Similarly, the quantum radar or sensing concept utilizing entangled particles is particularly interesting. Quantum processing is still in its early stages and is being considered for beyond 6G visions, but progress in this area should be monitored in order to incorporate this technology into the \ac{GREM} framework. } 

\vspace{-10pt}
\subsection*{Standardization Efforts}
The potential application of wireless (\ac{WLAN} in particular) sensing has led to interest from standardization bodies such as IEEE 802.11, resulting in the formation of a study group. Recent developments indicate its imminent evolution to a task group, targeted at modifying the \ac{PHY} and \ac{MAC} layers of the 802.11 standard to support sensing operation in 1-7.125 GHz and above 45 GHz frequency bands
. A particularly interesting topic of discussion in this regard is the development of channel models specifically for sensing applications \cite{zhang2020discussion}. Even if not feasible from the standardization perspective, this idea can motivate customized sensing models for different applications.

\vspace{-5.5pt}
\section*{Conclusion}
\label{Sec:Conclusion}
\par{With the successful trial deployments of \ac{5G} systems in various parts of the world, the attention of industry and academia has turned towards the development of the vision of 6G and identification of its enabling technologies. One common theme among the majority of these works is the need for an intelligent, flexible, efficient, and secure network. This work discusses the \ac{GREM} concept - comprising of but not limited to, knowledge regarding network infrastructure, propagation environment, attributes of the physical signal (multiple accessing scheme, waveform, modulation, etc.), and characteristics of the user equipment - as an enabler of the future wireless networks. The authors envision this \ac{GREM} to lend intelligence to the network by assisting in the improved deployment of nodes, better resource allocation, enhanced sensing and efficient communication.}

\section*{Acknowledgment}
The work of H. Arslan was supported by the Scientific and Technological Research Council of Turkey (TUBITAK) under Grant No. 5200030.

\vspace{-5pt}

\vspace{-170pt}
\begin{IEEEbiographynophoto}{Halise~T{\"u}rkmen} received her B.S and M.Sc degrees in mechatronics engineering from Marmara University and Istanbul Technical University in 2016 and 2019, respectively. Currently she is pursuing her Ph.D. degree at Istanbul Medipol University, Turkey. Her research interests include radio environment monitoring and sensing for enabling 5G and 6G systems.
\end{IEEEbiographynophoto}

\vspace{-170pt}
\begin{IEEEbiographynophoto}{Muhammad Sohaib J. Solaija}[S'16] received his B.E and M.Sc degrees in electrical engineering from National University of Science and Technology, Islamabad, Pakistan in 2014 and 2017, respectively. Currently he is pursuing his Ph.D. degree at Istanbul Medipol University, Turkey. His research focuses on interference modeling and coordinated multipoint (CoMP) implementation for 5G and beyond wireless systems.
\end{IEEEbiographynophoto}

\vspace{-170pt}
\begin{IEEEbiographynophoto}{Haji M. Furqan} received his B.E and M.Sc degrees in electrical engineering from COMSATS Institute of Information Technology (CIIT), Islamabad, Pakistan in 2012 and 2014, respectively. He received his PhD degree from Istanbul Medipol University, Turkey, where he is currently a Post-Doc researcher. His research focuses on physical layer security, cooperative communication, adaptive index modulation, OFDM, V2X, cryptography, 5G systems, and wireless channel modeling and characterization.
\end{IEEEbiographynophoto}

\vspace{-170pt}
\begin{IEEEbiographynophoto}{H{\"u}seyin Arslan}[S'95--M'98--SM'04--F'15] received his B.S. degree in electrical and electronics engineering from Middle East Technical University in 1992, and his M.S. and PhD degrees in electrical engineering from Southern Methodist University, Dallas, Texas, in 1994 and 1998, respectively. He is a professor of electrical engineering at the University of South Florida and the Dean of the School of Engineering and Natural Sciences at Istanbul Medipol University, Turkey. His current research interests are on physical layer security, mmWave communications, small cells, multi-carrier wireless technologies, co-existence issues on heterogeneous networks, aeronautical (high altitude platform) communications and \textit{in vivo} channel modeling, and system design. 
\end{IEEEbiographynophoto}

\end{document}